\documentstyle[aps,preprint,amssymb,epsfig]{revtex}

\begin{document}


\def\calA{{\cal A}}
\def\calH{{\cal H}}
\def\calL{{\cal L}}
\def\calO{{\cal O}}

\def\vecp{{\vec p}}

\def\etal{{\it et al.}}
\def\ibid#1#2#3{{\it ibid}. {\bf #1}, #3 (#2)}

\def\epjc#1#2#3{Eur. Phys. J. C {\bf #1}, #3 (#2)}
\def\ijmpa#1#2#3{Int. J. Mod. Phys. A {\bf #1}, #3 (#2)}
\def\jhep#1#2#3{J. High Energy Phys. {\bf #1}, #3 (#2)}
\def\mpl#1#2#3{Mod. Phys. Lett. A {\bf #1}, #3 (#2)}
\def\npb#1#2#3{Nucl. Phys. {\bf B#1}, #3 (#2)}
\def\plb#1#2#3{Phys. Lett. B {\bf #1}, #3 (#2)}
\def\prd#1#2#3{Phys. Rev. D {\bf #1}, #3 (#2)}
\def\prl#1#2#3{Phys. Rev. Lett. {\bf #1}, #3 (#2)}
\def\rep#1#2#3{Phys. Rep. {\bf #1}, #3 (#2)}
\def\zpc#1#2#3{Z. Phys. {\bf #1}, #3 (#2)}


\title{Photon polarization in $B\to K_{\rm res}\gamma$
\footnote{Contribution to the Proceedings of the 2nd International Conference
on Flavor Physics (ICFP 2003), KIAS, Seoul, Korea, 6-11 Oct. 2003}}
\author{Jong-Phil Lee}
\address{Department of Physics and IPAP, Yonsei University, Seoul, 120-749, Korea\\{\rm e-mail: jplee@phya.yonsei.ac.kr}}

\tighten
\maketitle

\begin{abstract}
In this talk, I introduce recently proposed method for measuring the photon 
polarization in $B\to K_{\rm res}\gamma$.
It is also shown that anomalous right-handed top couplings can severely
affect the photon polarization without spoiling the well measured branching
ratio of $B\to X_s\gamma$.
Estimated number of $B$ mesons needed for measuring the photon polarization is
already within the reach of current $B$ factories.
\end{abstract}
\pagebreak


Radiative $B$ decay of $B\to X_s\gamma$ is very important in testing the
standard model (SM) and constraining the new physics.
So far the inclusive and exclusive branching ratios as well as polarization of
the emitted photon have been extensively studied
\cite{Song,Atwood,Mannel,Melikhov}.
In the SM, photons from $b\to s\gamma$ are expected to be predominantly left 
handed up to $\calO(m_s/m_b)$.
Recently, Gronau {\it et al.} proposed a new way of measuring the photon
polarization in $B\to K_{\rm res}\gamma$ \cite{Gronau}.
Several Kaon resonances and some of their properties are listed in Table 1.
\par
\begin{figure}
\begin{center}
\epsfig{file=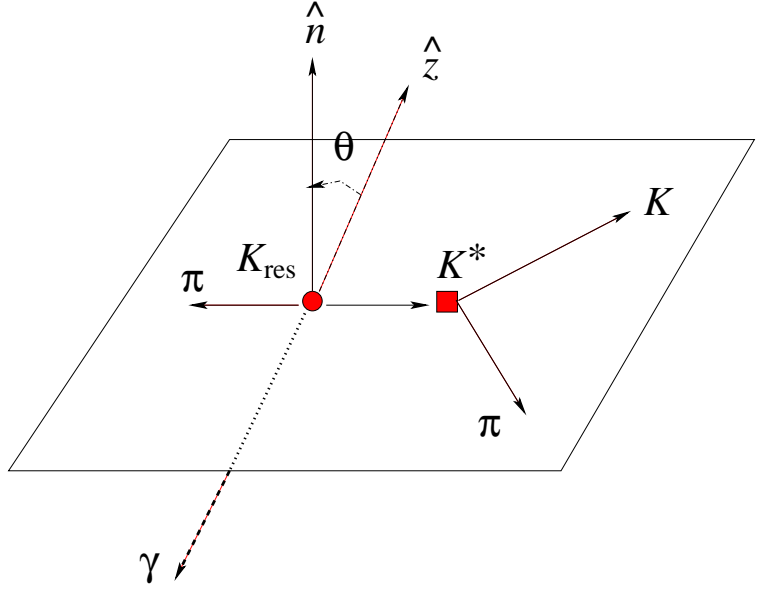,height=8cm}
\end{center}
\caption{Decay plane of $K_{\rm res}\to K\pi\pi$ in the $K_{\rm res}$-rest 
frame}
\label{decayplane}
\end{figure}

In this approach, the photon polarization is proportional to the integrated
up-down asymmetry of the emitted photon.
The asymmetry can be constructed through the triple vector product
$\vecp_\gamma\cdot(\vecp_1\times\vecp_2)$, where $\vecp_\gamma$ is the photon 
momentum and $\vecp_1$, $\vecp_2$ are two of the daughter hadron momenta, 
measured in the $K_{\rm res}$ rest frame.
Thus the three body decay of $K_{\rm res}$ is essential in this analysis.
The situation is depicted in Fig.\ \ref{decayplane}.
One of the remarkable result is that the so called photon polarization 
parameter $\lambda_\gamma$, which encodes the left-right asymmetry of the 
emitted photon polarization, is universally determined by the Wilson 
coefficients of the effective Hamiltonian.
Let me just show this point briefly.
\par
In the radiative $B$ decay of ${\bar B}\to{\bar K}_{\rm res}^{(i)}\gamma$,
the photon polarization in this process is naturally defined as follows:
\begin{equation}
\lambda_\gamma^{(i)}=
\frac{|A_R^{(i)}|^2-|A_L^{(i)}|^2}{|A_R^{(i)}|^2+|A_L^{(i)}|^2}~,
\end{equation}
where 
$A^{(i)}_{L(R)}\equiv\calA({\bar B}\to{\bar K}_{\rm res}^{(i)}\gamma_{L(R)})$
is the weak amplitude for left(right)-polarized photon.
The authors of \cite{Gronau} simply argued that 
\begin{equation}
\langle K_{\rm res}^{(i)R}\gamma_R|O_{12}'|{\bar B}\rangle=(-1)^{J_i-1}P_i
\langle K_{\rm res}^{(i)L}\gamma_L|O_{12}|{\bar B}\rangle~,
\label{LR}
\end{equation}
where $J_i (P_i)$ is the resonance spin (parity).
The usual electromagnetic operator $O_{12}$ and its chiral conjugate $O_{12}'$
are defined below.
The relation \ref{LR} implies that 
$|A_R^{(i)}|/|A_L^{(i)}|=|C_{12}'|/|C_{12}|$, and further
\begin{equation}
\lambda_\gamma^{(i)}=\frac{|C_{12}'|^2-|C_{12}|^2}{|C_{12}'|^2+|C_{12}|^2}
\equiv\lambda_\gamma~.
\label{lambdagamma}
\end{equation}
Here $C_{12}^{(\prime)}$ is the Wilson coefficient corresponding to the
operator $O_{12}^{(\prime)}$.
Thus the photon polarization parameter $\lambda_\gamma$ is independent of the
$K_{\rm res}$ states and is universally determined by the Wilson coefficients.
It is also free from the hadronic uncertainty which usually originates from the
weak form factors.
The relation (\ref{LR}) ensures that the common form factors are involved in
$\calA_L$ and $\calA_R$, being canceled in the ratio.
In the SM, $\lambda_\gamma\approx -1$ ($+1$ for ${\bar b}\to{\bar s}\gamma$) 
since the left-handedness is dominant, $|C_{12}'|/|C_{12}|\approx m_s/m_b$.
\par
To relate $\lambda_\gamma$ with the physical observable, a full analysis
including the strong decays of $K_{\rm res}$ must be implemented.
As mentioned before, there needs at least three hadrons in the final state
to see the up-down asymmetry.
One can readily calculate the angular distribution of 
${\bar B}\to{\bar K}\pi\pi$ \cite{Gronau}, and find that terms containing
the photon polarization parameter $\lambda_\gamma$ are proportional to the
up-down asymmetry of the emitted photon momentum with respect to the
$K\pi\pi$ decay plane.
The authors of \cite{Gronau} predicted that the integrated up-down asymmetry
in $K_1(1400)\to K\pi\pi$ is $(0.33\pm0.05)\lambda_\gamma$, which is quite
large compared to other $K_{\rm res}$ decays.
Estimated number of $B{\bar B}$ pairs required to measure the asymmetry is 
about $10^8$, which is already within the reach of current $B$ factories.
A stringent test of SM as well as constraints on new physics by the photon
polarization is therefore near at hand.
\par
In this talk, I investigate the effects of anomalous right-handed couplings
on the photon polarization parameter \cite{jplee}.
The left-right (LR) symmetric model and the minimal supersymmetric standard 
model (MSSM) can provide such new couplings.
The LR model is one of the natural extension of the SM, based on the 
$SU(2)_L\times SU(2)_R\times U(1)$ gauge group.
Besides the usual left-handed quark mixing, right-handed quark mixing is also
possible in the LR model.
Without a manifest symmetry between the left- and right-handed sectors, the
right-handed quark mixing is not necessarily the same as the left-handed quark
mixing governed by the CKM paradigm.
Consequently, there are additional right-handed charged current interactions 
with couplings different from the left ones, which are suppressed by the heavy
extra $W$ boson \cite{Babu}.
\par
In the unconstrained MSSM (uMSSM), the gluino-involved loop can contribute to 
the "wrong" chirality operator through the left-right squark mixing 
\cite{Everett}.
There can be a special case where $W$, Higgs, chargino, and gluino 
contributions to the ordinary Wilson coefficient $C_{12}$ tend to cancel each
other while the "wrong" chirality coefficient $C_{12}'$ gives the dominant
contribution.
\par
In the present analysis, however, I just concentrate on the anomalous 
right-handed top quark couplings ${\bar t}bW$ and ${\bar t}sW$, ignoring
the effects of additional left-handed interactions and new particles, and not
specifying the underlying models.
I introduce dimensionless parameters $\xi_s$ and $\xi_b$ where the anomalous
right-handed ${\bar t}sW$ and ${\bar t}bW$ couplings are encapsulated, 
respectively.
Up to the leading order of $\xi$, ordinary Wilson coefficients are modified
through the loop-function corrections proportional to $\xi_b$.
On the other hand, there appear new chiral-flipped operators in the effective 
Hamiltonian.
The corresponding new Wilson coefficients are proportional to $\xi_s$, with
new loop functions.
\par
It was recently shown that there is a parameter space of $(\xi_b, \xi_s)$ 
where the discrepancy of $\sin 2\beta$ between $B\to J/\psi K$ and $B\to\phi K$
is well explained while satisfying the $B\to X_s\gamma$ constraints \cite{KYL}.
Though the allowed values of $\xi$ are rather small, 
modified Wilson coefficients
of the colormagnetic and electromagnetic operators $O_{11}$, $O_{12}$ involve 
large enhancement factor of $m_t/m_b$.
Fortunately, the photon polarization parameter $\lambda_\gamma$ depends only on
the Wilson coefficient $C_{12}$ and its chiral-flipped partner $C'_{12}$, 
irrespective of the species of $K_{\rm res}$.
Thus the photon polarization parameter $\lambda_\gamma$, or the up-down
asymmetry of the emitted photon is quite sensitive to the new right-handed
couplings.
\par
The effective Lagrangian containing possible right-handed couplings can be
written as
\begin{equation}
\calL=-\frac{g}{\sqrt{2}}\sum_{q=s,b}V_{tq}
 {\bar t}\gamma^\mu(P_L+\xi_q P_R)q W^+_\mu+{\rm h.c.}~,
\end{equation}
where $P_{L,R}$ are the usual chiral projection operators.
With new dimensionless parameters $\xi_{b,s}$, the effective Hamiltonian for
the radiative $B$ decays has the form of
\begin{eqnarray}
\calH_{\rm rad}&=&-\frac{4G_F}{\sqrt{2}}V_{ts}^*V_{tb}\Big[
 C_{12}(\mu)O_{12}(\mu)+C'_{12}(\mu)O'_{12}(\mu)\Big]~,\nonumber\\
O_{12}&=&\frac{e}{16\pi^2}m_b{\bar s}P_R\sigma_{\mu\nu}bF^{\mu\nu}~,
\end{eqnarray}
and $O_{12}'$ is the chiral conjugate of $O_{12}$.
After matching at $\mu=m_W$, the Wilson coefficients are given by, in the SM,
\cite{Inami}
\begin{eqnarray}
C_{12}(m_W)&=&F(x_t)\nonumber\\
 &=&\frac{x_t(7-5x_t-8x_t^2)}{24(x_t-1)^3}
 -\frac{x_t^2(2-3x_t)}{4(x_t-1)^4}\ln x_t ~,\nonumber\\
C_{12}'(m_W)&=&0~,
\end{eqnarray}
where $x_t=m_t^2/m_W^2$.
Turning on the right-handed ${\bar t}bW$ and ${\bar t}sW$ couplings, the
Wilson coefficients are modified as 
\begin{eqnarray}
C_{12}(m_W)&\to&F(x_t)+\xi_b\frac{m_t}{m_b}F_R(x_t)~,\nonumber\\
C_{12}'(m_W)&\to&\xi_s\frac{m_t}{m_b}F_R(x_t)~,
\end{eqnarray}
with the new loop function \cite{KYL,Cho}
\begin{equation}
F_R(x)=\frac{-20+31x-5x^2}{12(x-1)^2}+\frac{x(2-3x)}{2(x-1)^3}\ln x~.
\end{equation}
Scaling down to $\mu=m_b$ is accomplished by the usual renormalization group
(RG) evolution.
Here the RG improved Wilson coefficients of \cite{KYL} are used.
\par
\begin{figure}
\begin{center}
\begin{tabular}{cc}
\epsfig{file=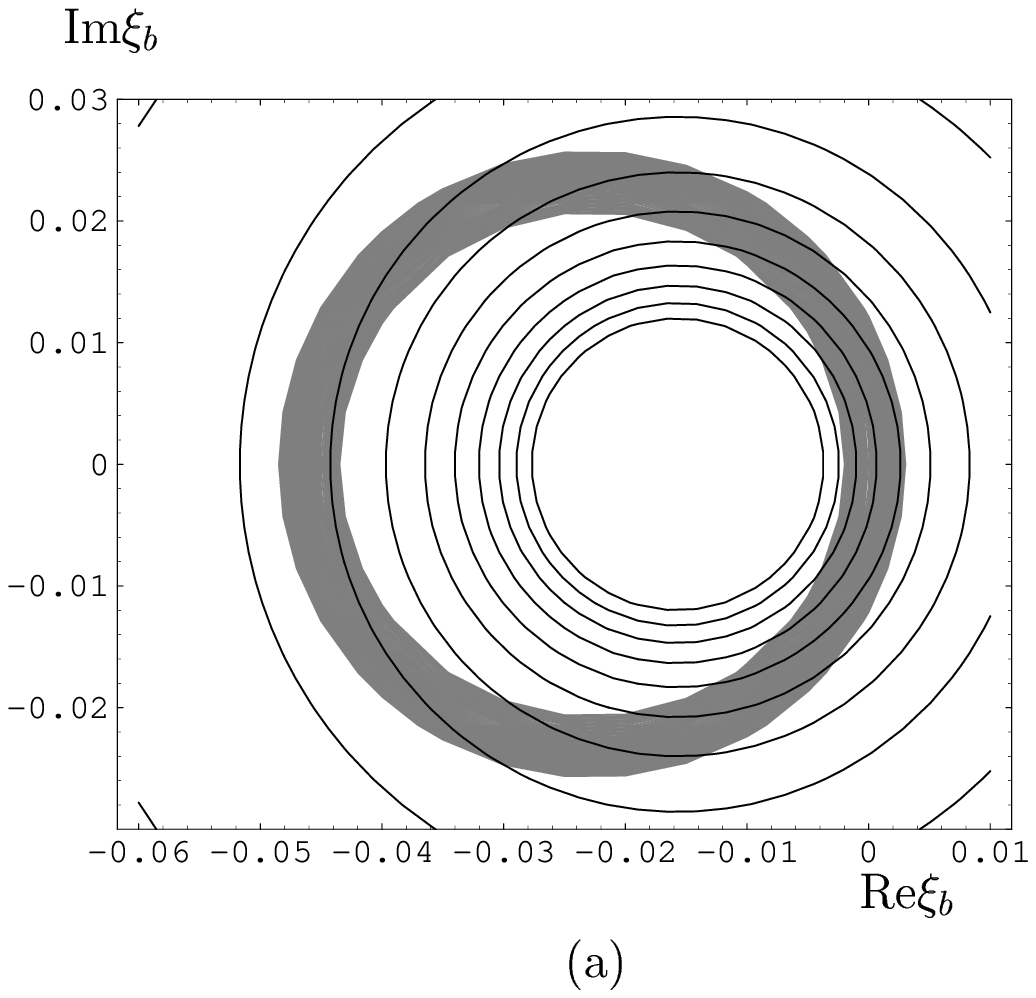,height=8cm}&
\epsfig{file=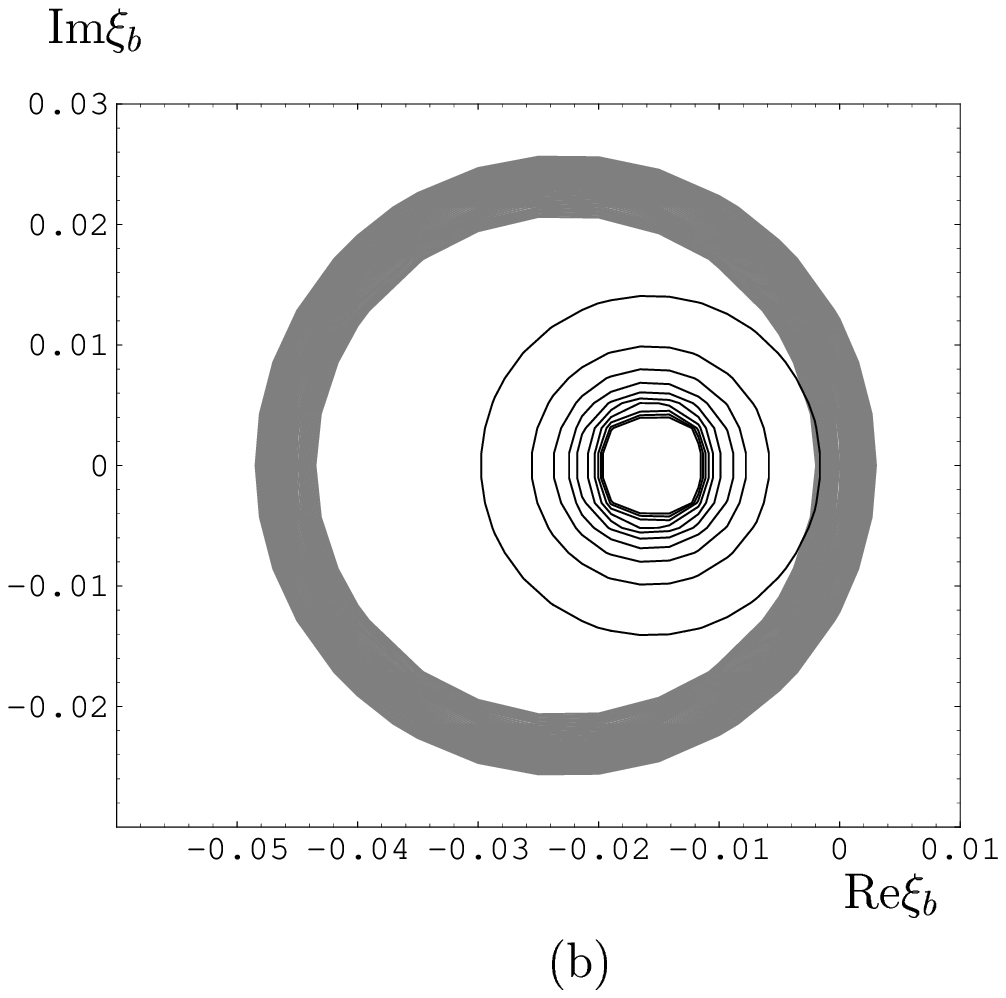,height=8cm}
\end{tabular}
\end{center}
\caption{Contour plot of $\lambda_\gamma$ for (a) $|\xi_s|=0.012$ and (b)
$|\xi_s|=0.001$. Concentric circles correspond to, from outside to inside, 
$\lambda_\gamma=-0.9$, $-0.8$, $\cdots$, $0$ in (a) and 
$\lambda_\gamma=-0.99$, $-0.98$, $\cdots$ ,$-0.9$ in (b), 
respectively.
Shaded ring is the allowed region from $B\to X_s\gamma$.}
\label{contour}
\end{figure}

Now let us examine the effects of newly introduced $\xi_{b,s}$ on 
$\lambda_\gamma$.
New couplings $\xi_{b,s}$ are strongly constrained by the branching ratio
${\rm Br}(B\to X_s\gamma)$ and the $CP$ asymmetry $A_{CP}(B\to X_s\gamma)$.
We use the weighted average of the branching ratio \cite{Everett,KYL}
\begin{equation}
{\rm Br}(B\to X_s\gamma)=(3.23\pm0.41)\times 10^{-4}~,
\end{equation}
from the measurements of Belle \cite{Belle}, CLEO \cite{CLEO}, and ALEPH 
\cite{ALEPH} groups.
The $CP$ violating asymmetry in $B\to X_s\gamma$, defined by
\begin{equation}
A_{CP}(B\to X_s\gamma)=
 \frac{\Gamma({\bar B}\to X_s\gamma)-\Gamma(B\to X_{\bar s}\gamma)}
 {\Gamma({\bar B}\to X_s\gamma)+\Gamma(B\to X_{\bar s}\gamma)}~,
\end{equation}
is measured by CLEO \cite{CLEO2}:
\begin{equation}
A_{CP}(B\to X_s \gamma)=(-0.079\pm0.108\pm 0.022)(1.0\pm 0.030)~.
\end{equation}
The explicit expressions of the branching ratio and the $CP$ asymmetry are 
given in \cite{Kagan} in terms of the evolved Wilson coefficients at the
$\mu=m_b$ scale.
We adopt the constraints on $\xi$ established in \cite{KYL} from these 
experimental and theoretical results at $2\sigma$ C.L.:
\begin{eqnarray}
-0.002&<&{\rm Re}\xi_b+22|\xi_b|^2<0.0033~,\nonumber\\
-0.299&<&\frac{0.27{\rm Im}\xi_b}{0.095+12.54{\rm Re}\xi_b+414.23|\xi_b|^2}
<0.141~,\nonumber\\
|\xi_s|&<&0.012~.
\label{constraints}
\end{eqnarray}
Figure \ref{contour} shows the contour plot of $\lambda_\gamma$ for various 
values of $\xi_{b,s}$.
Shaded ring denotes the allowed region by (\ref{constraints}).
Since the measured $CP$ asymmetry has rather large errors, constraints on 
$\xi$ are mainly from the ${\rm Br}(B\to X_s\gamma)$.
In Fig.\ \ref{contour}, two contours for small and large value of $\xi_s$ are 
shown as an illustration.
As one can expect from (\ref{lambdagamma}), deviation of $\lambda_\gamma$ from
$-1$ can be sizable if $\xi_s$ is large (Fig.\ \ref{contour} (a)).
On the other hand, if $\xi_s$ is very small, $\lambda_\gamma\approx -1$,
irrespective of $\xi_b$ (Fig.\ \ref{contour} (b)).
\par
\begin{figure}
\begin{center}
\epsfig{file=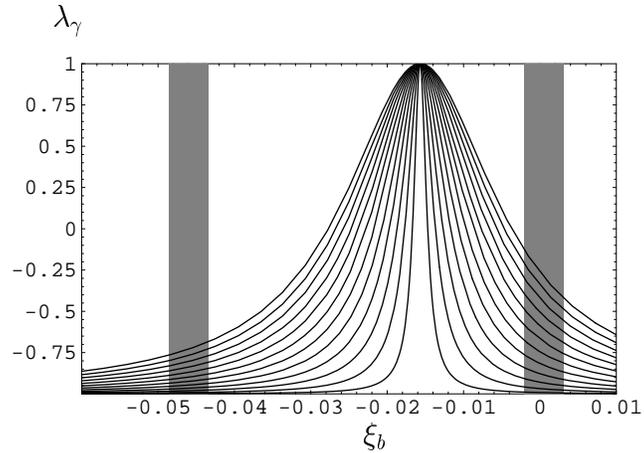,height=6cm}
\end{center}
\caption{Plots of $\lambda_\gamma$ as a function of $\xi_b$ for various $\xi_s$,
assuming that $\xi_{b,s}$ are real. 
Each curves corresponds to $\xi_s=0.001$, $0.002$, $\cdots$, $0.012$, from
bottom to top, respectively.
Shaded bands are the allowed region from $B\to X_s\gamma$.}
\label{real}
\end{figure}

As can be seen in Fig.\ \ref{contour} (a), large value of $\lambda_\gamma$ can 
be obtained in the region of ${\rm Im}\xi_b=0$.
In Fig.\ \ref{real}, we give plots of $\lambda_\gamma$ vs $\xi_b$ for 
different values of $\xi_s$.
Here we assumed that $\xi_{b,s}$ are all real for simplicity.
Shaded bands are the allowed region by (\ref{constraints}).
In this case, a large deviation of $\lambda_\gamma$ from $-1$ is possible in 
the right-hand-side band.
This is quite natural since one can expect large $\lambda_\gamma$ in the region
of large $\xi_s$ and small $\xi_b$, by the inspection of (\ref{lambdagamma}).
We have
\begin{equation}
-1\le\lambda_\gamma\lesssim -0.12~.
\label{polbound}
\end{equation}
It should be noticed that current experimental bounds on $B\to X_s\gamma$ do 
not allow the different sign of $\lambda_\gamma$ compared to the SM prediction
at $2\sigma$ level.
Note that the upper bound of $\lambda_\gamma$ is chosen at the edge point 
of the allowed parameter space, say, $(\xi_b,\xi_s)=(-0.0021,0.012)$. 
If the new couplings are flavor-blind, i.e. $\xi_b=\xi_s$, then 
$\lambda_\gamma\simeq -0.96$ for $\xi_{b,s}=-0.002$.
Thus a large amount of reduction in $|\lambda_\gamma|$ implies that the 
anomalous right-handed couplings are flavor dependent.
\par
It is quite interesting to compare our result with that of the uMSSM 
\cite{Everett}.
The authors of Ref.\ \cite{Everett} proposed the "$C_{12}'$-dominated" scenario,
where the total contribution to $C_{12}$ is negligible while the main 
contribution to the ${\rm Br}(b\to s\gamma)$ is given by $C_{12}'$.
This is possible when the chargino, neutralino, and gluino contributions to
$C_{12}$ are canceled out by the $W$ and Higgs contributions.
Now that the size of $C_{12}$ is very small, they expect 
$\lambda_\gamma\approx +1$ as an extreme case, quite contrary to the SM 
predictions.
This is also very distinguishable from our result, since (\ref{polbound}) does
not allow the sign flip of $\lambda_\gamma$.
Thus the sign of $\lambda_\gamma$ is a very important landmark indicating
which kind of new physics is involved, if exists.
In case of ${\rm sgn}(\lambda_\gamma)>0$, models which produce only the 
anomalous right-hand top vertices would be disfavored.
At least we might need new particles, or new mechanism to make the $C_{12}'$
dominant.
\par
As introduced earlier, the integrated up-down asymmetry of 
$K_1(1400)(\to K^*\pi, \rho K)\to K^0\pi^+\pi^0$ or $K^+\pi^-\pi^0$ is reported 
to be $(0.33\pm0.05)\lambda_\gamma$ \cite{Gronau}, where
the uncertainty is a combined one from the uncertainties of $D$- and $S$-wave
amplitudes in the $K^*\pi$ channel, and $\rho K$ amplitude.
Within the SM where $\lambda_\gamma\approx -1$, it means that about 80
charged and neutral $B$ and ${\bar B}$ decays into $K\pi\pi\gamma$ are needed
to measure an asymmetry of $-0.33$ at $3\sigma$ level.
The authors of \cite{Gronau} estimated that at least $2\times 10^7~B{\bar B}$
pairs of both neutral and charged are required, with the use of 
${\rm Br}(B\to K_1(1400)\gamma)=0.7\times 10^{-5}$ and 
${\rm Br}(K_1(1400)\to K^*\pi)=0.94\pm 0.06$.
\par
If the anomalous right-handed couplings were present, we would need more 
$B{\bar B}$ pairs because new couplings will reduce the value of 
$|\lambda_\gamma|$.
For example, in case of $\lambda_\gamma=-0.5$, we need 4 times larger number
of $B{\bar B}$ pairs ($8\times 10^7$) to see the $3\sigma$ deviation of the
up-down asymmetry from both zero and the SM prediction.
Fortunately, this is already within the reach of current $B$ factories 
\cite{BABAR}.
If the deficiency of the up-down asymmetry were found, then the direct searches
of the anomalous top couplings at the LHC in coming years would be very 
exciting to check the consistency.
\par
In summary, we have analyzed the effects of anomalous right-handed top
couplings on the photon polarization in radiative $B\to K_{\rm res}\gamma$ 
decays.
The photon polarization parameter $\lambda_\gamma$ defined by the ratio of
the relevant Wilson coefficients is a useful observable for measuring the photon
helicity.
It is found that the new couplings can reduce $|\lambda_\gamma|$ significantly,
compared to the SM prediction of $\lambda_\gamma\approx -1$,
while satisfying the strong constraints from the measured branching ratio and 
$CP$ asymmetry of $B\to X_s\gamma$.
We also find that the anomalous right-handed top couplings would not produce 
different sign of $\lambda_\gamma$ from the SM prediction.
This is a crucial point to distinguish our case from other scenarios such as 
uMSSM where an extreme value of $\lambda_\gamma=+1$ can be possible through
the "$C_{12}'$-dominated" mechanism.
The importance of present work also lies in the fact that current $B$ factories 
can produce enough $B{\bar B}$ pairs to analyze the photon polarization in
$B\to K_{\rm res}\gamma$.
\par
This work was supported by the BK21 program of the Korean Ministry of Education.



\begin{table}
\begin{tabular}{ccccc}
Resonances&$J^P$&$(M_{\rm res}, \Gamma_{\rm res})$ (MeV)&Decay Mode&Br (\%)\\
\hline
$K_1(1270)$&$1^+$&$(1273\pm7, 90\pm20)$&$\rho K$&$42\pm6$\\
&&& $K^*\pi$ & $16\pm5$\\
&&& $K^{*0}(1430)\pi$ & $28\pm4$\\
$K_1(1400)$&$1^+$&$(1402\pm7, 174\pm13)$&$K^*\pi$&$94\pm6$\\
&&& $\rho K$ & $3.0\pm3.0$\\
$K^*(1410)$&$1^-$&$(1414\pm15, 232\pm21)$&$K^*\pi$&$>40$\\
&&& $\rho K$ & $<7$\\
$K_2^*(1430)$&$2^+$&$(1425.6\pm1.5, 98.5\pm2.7)$&$K^*\pi$&$24.7\pm1.5$\\
&& (charged $K_2^*$) & $\rho K$ & $8.7\pm0.8$
\end{tabular}
\caption{Some Kaon resonances around $1300\sim 1400$ MeV [5].}
\end{table}

\end{document}